# Development of a Transferable Reactive Force Field of P/H Systems: Application to the Chemical and Mechanical Properties of Phosphorene

Hang Xiao[1], Xiaoyang Shi[1], Feng Hao[1], Xiangbiao Liao[1], Yayun Zhang[1,3] and Xi Chen[1,2]

[1] *Columbia Nanomechanics Research Center, Department of Earth and Environmental Engineering, Columbia University, New York, NY 10027, USA*

[2] *School of Chemical Engineering, Northwest University, Xi'an 710069, China*

[3] *College of Power Engineering, Chongqing University, Chongqing 400030, China*

## Abstract

We developed ReaxFF parameters for phosphorus and hydrogen to give a good description of the chemical and mechanical properties of pristine and defected black phosphorene. ReaxFF for P/H is transferable to a wide range of phosphorus and hydrogen containing systems including bulk black phosphorus, blue phosphorene, edge-hydrogenated phosphorene, phosphorus clusters and phosphorus hydride molecules. The potential parameters were obtained by conducting global optimization with respect to a set of reference data generated by extensive *ab initio* calculations. We extended ReaxFF by adding a 60 ° correction term which significantly improved the description of phosphorus clusters. Emphasis was placed on the mechanical response of black phosphorene with different types of defects. Compared to the nonreactive SW potential,[1] ReaxFF for P/H systems provides a significant improvement in describing the mechanical properties of the pristine and defected black phosphorene, as well as the thermal stability of phosphorene nanotubes. A counterintuitive phenomenon is observed that single vacancies weaken the black phosphorene more than double vacancies with higher formation energy. Our results also showed that the mechanical response of black phosphorene is more sensitive to defects in the zigzag direction than that in the armchair direction. Since ReaxFF allows straightforward extensions to the heterogeneous systems, such as oxides, nitrides, the proposed ReaxFF parameters for P/H systems also underpins the reactive force field description of heterogeneous P systems, including P-containing 2D van der Waals heterostructures, oxides, etc.

# 1. Introduction

In recent years, two-dimensional (2D) materials have attracted much interest because of their fascinating electronic,[2,3] mechanical,[4] optoelectronic,[5,6] and chemical [7,8] properties. The epic discovery of graphene opened up the possibility of isolating and studying the intriguing properties of a whole family of 2D materials including the 2D insulator boron nitride (BN),[9–11] 2D semiconductor molybdenum disulfide [9,12,13] and recently, 2D phosphorus, i.e. phosphorene.[14,15] Single layer black phosphorus, i.e. phosphorene, was obtained in experiments in 2014.[15] Because of its tunable band gap and high carrier mobility, phosphorene holds great potential in electronic and optoelectronic applications.

Over the past decade, tremendous success has been achieved in the synthesis of 2D materials. However, the cycles of synthesis, characterization and test for 2D materials are slow and costly, which inspired the development of computational tools to design new 2D materials [16–18] and to provide guidance for the fabrication of 2D devices.[19–22] Although *ab initio* methods (such as density functional theory, DFT) provide accurate description of the electronic structure of 2D crystals, they are limited to small systems (several hundreds of atoms) with short time scales (picoseconds). To the contrary, molecular dynamics simulations powered by force fields are able to reach much larger scale with much longer time. To date, several force fields have been developed for black phosphorus.

A valence force field (VFF) for black phosphorus was first proposed in 1982 [23] and used to study the elastic properties in black phosphorus. More recently, Jiang *et al*.[1] developed a Stillinger-Weber (SW) potential for phosphorene based on the VFF model by fitting parameters to experimental phonon spectrum. In the SW potential, the energy parameters were taken from the VFF model, and geometrical parameters were derived analytically from the equilibrium state of individual potential terms. While both VFF model and SW potential have been used to describe phonons and elastic deformations, they are not suitable to describe states far from equilibrium.[24] Moreover, the SW potential strongly underestimated the Young's modulus of black phosphorene in the zigzag direction.[25] Due to its nonreactive nature, SW potential also had difficulty describing phosphorene with defects. An improved force field which balances accuracy and computational efficiency is therefore highly desirable. In 2001, van Duin *et al*. developed a reactive force field (ReaxFF) for hydrocarbons.[26] ReaxFF is a bond order interaction model, capable of handling bond breaking and forming with associated changes in atomic hybridization. Since its development, ReaxFF model has been applied to a wide range of systems.[27–33] To the best of our knowledge, a ReaxFF model for phosphorene system is still lacking.

In this work, we develop a ReaxFF parameter set for P and H to describe the chemical and mechanical properties of pristine and defected black phosphorene. ReaxFF for P/H is transferable to a wide range of phosphorus and hydrogen containing systems including bulk black phosphorus, blue phosphorene, hydrogenated phosphorene, phosphorus clusters and phosphorus hydride molecules. The ReaxFF parameters for P/H were fitted to a set of reference data generated by extensive *ab initio* calculations. The proposed ReaxFF for P/H provides a distinctive improvement in describing the thermomechanical properties the pristine and defected black phosphorene, as well as that of the phosphorene nanotubes (PNTs) over the SW potential. The ReaxFF parameters for P/H presented here provide a first step in the development of a reactive force field description for the heterogeneous P systems.

# 2. Methodology

## 2.1 DFT Calculations

The fitting data used for P/H systems was obtained from DFT calculations performed with the Cambridge series of total-energy package (CASTEP).[34,35] For these calculations, ultrasoft pseudopotentials were used to describe the core electrons and the electron exchange-correlation effects was described by the Perdew–Burke–Ernzerhof (PBE) [36] generalized gradient approximation. In this work, the empirical dispersion correction schemes proposed by Grimme (D2) [37] was used in combination with PBE functional. In computing the energies of phosphorus clusters, phosphorus hydride molecules and phosphorene with defects and adatoms, spin polarization was used to account for the energy contributions from magnetization. Periodic boundary conditions were used for all the calculations, with monolayer structures represented by a periodic array of slabs separated by a 15 Å thick vacuum region. A large 5 ×7 supercell of black phosphorene was adopted to study the effect of defects and adatoms. A plane wave cutoff of 520 eV was used to determine the self-consistent charge density. For condensed phases, Brillouin zone integrations were performed with Monkhorst-Pack[38] mesh with 0.02 Å$^{-1}$ $k$-point spacing. For cluster calculations, a cubic supercell of 20 Å (to ensure the interactions between clusters in adjacent cells is negligible) was used with the clusters or molecules placed at the center of the cell with Brillouin zone sampled at the Γ point. All geometries were optimized by CASTEP using the conjugate gradient method (CG) with convergence tolerances of a total energy within $5.0 \times 10^{-6}$ eV atom$^{-1}$, maximum Hellmann–Feynman force within 0.01 eV Å$^{-1}$, maximum ionic displacement within $5.0 \times 10^{-5}$ Å, and maximum stress within 0.01 GPa. For black phosphorene, the stress-strain responses in the armchair and zigzag directions were calculated using the method described in the references [39,40] with CASTEP. The CASTEP calculations showed good agreement with previous theoretical values for a variety of phosphorene properties: lattice constants,[15] Young's modulus in the armchair and zigzag directions,[41] formation energies of defects [42] and adatoms [43]. And the calculated lattice constants of bulk black phosphorus agreed well with experimental values.[44]

## 2.2 ReaxFF

The ReaxFF model [26–28] is a bond-order interaction model. For ReaxFF, the interatomic potential describes chemical reactions through a bond-order framework, in which the bond order is directly calculated from interatomic distances. Within the bond order framework, the electronic interactions (i.e. the driving force of the chemical bonding) are treated implicitly, allowing the method to simulate chemical reactions without expensive quantum chemical calculations. Typical empirical force field (EFF) potentials adopt empirical equations to describe the bond stretching, bond bending, and bond torsion events, with additional expressions to handle the van der Waals (vdW) and Coulomb interactions. These EFF potentials require a user-specified connectivity table, while ReaxFF is able to calculate the atom connectivity on the fly, which distinguishes ReaxFF from conventional EFF potentials since the breaking and forming of bonds can be captured during MD simulations.

For a ReaxFF description of P/H systems, the bond energies ($E_{bond}$) are corrected with over-coordination penalty energies ($E_{over}$). Energy contributions from valence angle ($E_{val}$) and torsion angle ($E_{tor}$) are included. Dispersion interactions are described by the combination of the original van der Waals term ($E_{vdw}$) and low-gradient vdW correction term ($E_{lgvdw}$) [31]. Energy contribution from Coulomb interactions ($E_{Coulomb}$) are taken into account between all atom pairs, where the atomic charges are calculated based on connectivity and geometry using the Electron Equilibration Method (EEM).[45] All energy terms except the last three are bond-order dependent and a detailed description of them (except $E_{60cor}$) can be found in Refs. [26,27,31]. The total energy is the summation of these energy pieces, shown by

$$E_{system} = E_{bond} + E_{over} + E_{val} + E_{60cor} + E_{tors} + E_{vdw} + E_{lgvdw} + E_{Coulomb} \quad (1)$$

The stability of $P_4$ cluster and the instability of larger phosphorus clusters has been an ongoing puzzle for several decades.[46] The phosphorus is often expected to favor valence angles near 101°.[47] If this is true, the strain energy of bonds in $P_4$ cluster (with 60° valence angles) should make it unstable. In QM calculations, this problem was resolved by including the effect of d-orbitals.[48] In order to address the stability of the $P_4$ cluster (and other phosphorus clusters with valence angles near 60°), we added a 60° angle correction term to Eq. 1[*].

$$E_{60cor} = -p_{cor1} \cdot f_1(BO_{ij}) \cdot f_2(BO_{jk}) \cdot \exp\left[-p_{cor2} * (\Theta_{60} - \Theta_{ijk})^2\right] \quad (2a)$$

$$f_1(BO_{ij}) = 1 - \exp(-p_{val3} \cdot BO_{ij}^{p_{cor3}}) \quad (2b)$$

---

[*] Note that, for the simulation of P/H systems with the 60° angle correction, one needs to use the force field file with 60° angle correction and recompile the LAMMPS package with our modified source file, *reaxc_valence_angles.cpp*. We verified that the 60° angle correction term would only affect the properties of P/H systems with valence angles near 60°. Therefore, for the simulation of condensed phases (either pristine or defected) and phosphorus hydride molecules, the original software of LAMMPS package can be used with the force field file without 60° angle correction. Because the lgvdw term was included in the ReaxFF, the "pair_style" command in the input file of LAMMPS should be: pair_style    reax/c NULL lgvdw yes.

$$f_2(BO_{jk}) = 1 - \exp(-p_{val3} \cdot BO_{jk}^{p_{cor3}}) \qquad (2c)$$

In section 3.2.2, it is demonstrated that the accuracy of cluster formation energies was significantly improved by the addition of the 60 ° angle correction term.

LAMMPS code [49] was used to perform MD calculations for the tensile behavior for the black phosphorene of dimension 27.5 ×25.8 Å at 1.0 K and 300.0 K. Periodic boundary conditions were employed in both the zigzag and armchair directions. The equation of motion was solved with a velocity Verlet algorithm, using a time step of 1.0 fs, which led to stable dynamics trajectories. The system was thermalized to steady state with the NPT (constant number of particles, constant pressure, and constant temperature) ensemble for 50 ps by the Nosé-Hoover [50,51] thermostat. Subsequently, the black phosphorene was stretched in zigzag or armchair direction at a strain rate of $10^9$ s$^{-1}$, and the stress in the lateral direction was fully relaxed. In computing the stress, the inter-layer spacing of 5.24 Å was used as the thickness of the black phosphorene. The Young's modulus was calculated from the stress-strain curve in the strain range [0, 0.01]. Following the same procedure of calculating the stress-strain curve for the defect-free black phosphorene, the MD calculations for defected phosphorene under tensile strain were conducted for the black phosphorene of dimension 27.5 ×25.8 Å at 1.0 K with one defect (in the form of single vacancy, double vacancy or Stone-Wales defect). For the stability analysis of PNTs, each PNT with the length of 10 supercells is equilibrated to a thermally stable state under NPT ensemble at a given temperature (from 0-800 K).

## 3. Results and Discussions

### 3.1 DFT Training of Force Field

The ReaxFF parameters for P/H systems were optimized using a modified version of the evolutionary algorithms (EA) software suite OGOLEM,[52,53] which is able to globally optimize ReaxFF parameter sets with high parallel efficiency. Based on DFT calculations for bulk black phosphorus, pristine and defected black phosphorene, blue phosphorene, phosphorus hydride molecules and phosphorus clusters, ReaxFF parameters were generated for P-P and P-H bond energies, P-P-P, H-P-P and H-P-H valence angle energies and for H-P-P-P and H-P-P-H torsion energies.

The parametrization of ReaxFF for P/H systems consisted of following steps:

(i) Training set of DFT data points was built for crystals, clusters and phosphorus hydride molecules. For the crystal phases, the energy-volume relationship of the black phosphorus crystal and the energy-area relationship of both black and blue phosphorene were deduced. The bond dissociation profiles of P-P bonds in the $P_2H_4$ and $P_2H_2$ molecules, and of P-H bonds in the $PH_3$ molecules were included. Energy profiles for angle distortion of P-P-P in the $P_3H_5$ molecule, of H-P-P in the $P_2H_4$ molecule, and of H-P-H in the $PH_3$ molecule were added. In these energy profiles, only the lowest-energy states (singlet, triplet or quintet depending on geometry) were included. The Mulliken charges for the phosphorus hydride molecules were added to the training set. A minimum number of terms in Eq. 1 were selected (starting with $E_{bond}, E_{over}, E_{val}, E_{vdw}, E_{Coulomb}$). The parameters were fitted to the training set using OGOELM [52,53].

(ii) The torsion angle term ($E_{tor}$), low gradient correction term ($E_{lgvdw}$), and 60° angle correction term ($E_{60cor}$) were added to the total energy function to obtain a refined fit to the training set. Energy profiles for torsion distortion of H-P-P-H in the $P_2H_4$ molecule and of H-P-P-P in the $P_4H_2$ molecule were included. Energies and geometries of phosphorene with different types of defects were added to the training set.

(iii) The global optimized parameters were validated by the comparison of properties calculated by ReaxFF to experimental and DFT data.

## 3.2 Parameterization and Validation of ReaxFF

Our final fitted, global optimized ReaxFF for P/H systems is given in Tables 1-7. The potential form is given in Eq. 1 (a detailed description of all terms can be found in Refs. [26,27,31]).

Unless otherwise stated, all ReaxFF results in the following discussion refer to our global optimized ReaxFF parameter set.

### 3.2.1 Relative stabilities for bulk black phosphorus, black and blue phosphorene

For ReaxFF to accurately describe phosphorus in the condensed phase, descriptions for different crystalline phases should be included in the DFT training set. Relative stabilities of the black phosphorus crystal as a function of unit cell volumes and relative stabilities of the both black and blue phosphorene as a function of unit cell in-plane areas were calculated. In general, the ReaxFF model gives a good description of lattice parameters of all three crystal phases (see Table 8) and shows a good consistency of the crystal structures of these crystal phases (see Fig. 1).

In particular, the DFT and ReaxFF results of cohesive energies are compared to SW results and experimental data in Table 9. The equilibrium cohesive energy of bulk phosphorus used in the fitting procedure was the experimental data [54] of -3.26 eV rather than the value computed from DFT (-3.43 eV). ReaxFF predicts a black phosphorus cohesive energy of -2.91 eV. The cohesive energy of black phosphorene calculated by ReaxFF is -2.84 eV, which slightly underestimates the DFT result of -3.35 eV. Still, ReaxFF provides a much better prediction of cohesive energy of phosphorene than that of SW potential, which underestimates the cohesive energy of phosphorene by an order of magnitude. ReaxFF are able to correctly reproduce the relative order of stability of three crystal phases (shown in Table 9). In Fig. 2(a) and Fig. 2(b), the results from ReaxFF correctly describe the relative stabilities of bulk black phosphorus for a broad range of cell volume, as well as that of black phosphorene for a broad range of cell area. In the training set, not all the data can be fitted equally well. For blue phosphorene (Fig. 2(c)), ReaxFF slightly overestimates the in-plane area of the unit cell, leading to a small offset of the energy profile of the relative stability. Given that no existing force field can describe the properties of blue phosphorene, the present ReaxFF may represent a major step forward.

### 3.2.2 Relative stabilities of phosphorus clusters

For ReaxFF to provide accurate description of phosphorus in clusters, the geometries and formation energies of P clusters of sizes 4,5,6 and 8 atoms are included in the training set. The formation energies per atom of clusters, $E_{cf}$, defined by

$$E_{cf} = E_c/n - E_{cohesive}(\text{bulk}) \qquad (3)$$

where $E_c$ is the energy of the relaxed phosphorus cluster with *n* atoms, $E_{cohesive}(\text{bulk})$ is the cohesive energy of the bulk black phosphorus. As can be seen from Fig. 3, ReaxFF is capable of providing a very good description of the geometries of P clusters. Table 10 shows that the cluster formation energies per atom calculated by ReaxFF with 60 °correction agree well with the DFT results. It is intriguing that a simple 60 °angle correction term is able to provide such a notable improvement in terms of cluster formation energies.

### 3.2.3 Potential energy curves for phosphorus hydride molecules

Data for selected phosphorus hydride molecules was also included in the training set to train the P-H interactions and to enhance the transferability of the ReaxFF for P/H systems. To include DFT data for P-H, P-H bonds, dissociation profiles were determined from DFT calculations for phosphine, $P_2H_2$ and $P_2H_4$ molecules. The bond dissociation profiles were generated from the equilibrium geometries of these molecules by changing the bond length from the equilibrium value while allowing other structural parameters to relax, which are shown in Fig. 4 (a-c). Only the lowest-energy states (singlet, triplet or quintet depending on geometry) were included in bond dissociation profiles. The DFT and ReaxFF curves are shown in Fig. 4 (a-c).

To include DFT data for P-P-P, H-P-P and H-P-H valence angles, $P_3H_5$, $P_2H_2$ and phosphine molecules were used, respectively. Following the same procedure of constructing the bond dissociation profiles, $P_3H_5$, $P_2H_2$ and phosphine molecules were geometry optimized to create reference states. Afterwards the valence angles were modified while other structural parameters were optimized. The resulting angle distortion curves are shown in Fig. 4 (d-f).

Energy profiles for torsion distortion of H-P-P-H in the $P_2H_4$ molecule and of H-P-P-P in the $P_4H_2$ molecule were also included in the training set. The torsion distortion curves were generated from the equilibrium geometries of these molecules by changing the relevant torsion angle from the equilibrium value while allowing other structural parameters to relax, which are shown in Fig. 4 (g-h).

In Fig. 4 (a, d, f, g, h), it is visible that the interactions between phosphorus and hydrogen atoms in phosphorus hydride molecules are well reproduced with ReaxFF. For the interactions between phosphorus atoms in phosphorus hydride molecules (see Fig. 4 (b, c, e)), agreement between the ReaxFF and DFT results are not perfect, because the crystal phases of phosphorus were prioritized over the phosphorus hydride molecules in ReaxFF. The depth of the ReaxFF potential well in Fig. 4(b) is shallow, in order to offset the errors in cohesive energy for bulk black phosphorus (cf. Table 9) and the ultimate strength of black phosphorene in zigzag direction (cf. Fig. 7).

### 3.2.4 Defects for black phosphorene

Properties and applications of 2D materials are strongly affected by defects,[55] which are generally induced by irradiations of ion or electron.[56] Defect engineering has emerged as an important approach to modulate the properties of 2D materials. Thus the accurate description of behavior of different types of defects in phosphorene is critical for ReaxFF of P/H systems. The structures and formation energies of single vacancy (SV), double vacancy (DV) and Stone-Wales (SW) defects are included in the training set. The defect formation energy, $E_{df}$, defined by

$$E_{df} = E_d - E_{cohesive}(\text{black}) \cdot n \tag{4}$$

where $E_d$ is the energy of the defected phosphorene (geometry optimized) with *n* phosphorus atoms, $E_{cohesive}(\text{black})$ is the energy per atom of the black phosphorene. Fig. 5 shows that ReaxFF performs very well in predicting defects geometries of all three types with respect to DFT calculations, whereas the SW potential fails to predict the structure of all three type of defects. From Table 11, ReaxFF provides a good description of the defect formation energy of single vacancy and double vacancy in phosphorene, as well as the relative stability between single vacancy and double vacancy. The formation energy of Stone-Wales defect is overestimated by 36% by ReaxFF. By comparison, for SW potential, the formation energies of single and double vacancy are seriously underestimated (see Table 11) and the Stone-Wales defect is unstable (see Fig. 5), leading to an erroneous 0 eV formation energy. Compared to SW potential, ReaxFF provides a significant improvement in describing different types of defects in phosphorene.

### 3.2.5 Adatoms for black phosphorene: a transferability test

Due to its 2D nature, the large surface area to volume ratio of a black phosphorene nanosheet leads to a high chemical activity to foreign atoms. Thus the accurate description of surface adatoms in phosphorene is an important objective for ReaxFF. Structures and formation energies of phosphorus and hydrogen adatoms for black phosphorene were withheld from the training set, to serve as the validation data. The adsorption energy of adatoms on phosphorene, $E_{ad}$, defined by

$$E_{ad} = E_{adsorp} - E_{psheet} - E_{atom} \qquad (5)$$

where $E_{adsorp}/E_{psheet}$ is the total energy of phosphorene with/without adatoms and $E_{atom}$ is the energy of the isolated atom. Fig. 6 shows that ReaxFF agrees very well with DFT calculations for predicting the adsorption structures of P and H adatoms. By contrast, the SW potential overestimates the bond length between P adatom and upper P atoms in black phosphorene. Without P-H interactions, SW potential is not capable to describe the H adatoms for black phosphorene. In Table 12, it can be seen that ReaxFF provides a good description of the binding energy of P adatom and slightly overestimates the binding energy of H adatom. However, the SW potential underestimates the binding energy of P adatom by an order of magnitude. Overall, ReaxFF provides a good description of P and H adatoms on black phosphorene. Since the structures and formation energies of P and H adatoms for black phosphorene were not included in the training set, these results indicate a good transferability of the ReaxFF for P/H systems.

## 3.3 Mechanical Property of Black Phosphorene Predicted by ReaxFF

In Table 13, the Young's modulus of black phosphorene in armchair and zigzag directions calculated by ReaxFF and SW potential are compared to DFT results. ReaxFF performs fairly well in reproducing the Young's modulus of black phosphorene in both directions, while SW potential underestimates the Young's modulus of black phosphorene in both directions. Fig. 7 (a) shows the stress-strain curves of black phosphorene in zigzag and armchair directions calculated with DFT, ReaxFF and SW potential. For zigzag direction, ReaxFF is able to capture the modulus change as the strain increases, providing a reasonable agreement in ultimate strength and failure strain. However, the SW potential severely underestimates the ultimate strength and failure strain in the zigzag direction. For armchair direction, ReaxFF overpredicts the failure strain while SW potential underpredicts it. The ultimate strength of black phosphorene in the armchair direction is slightly overestimated by ReaxFF, while it is severely underestimated by SW potential. ReaxFF yields a smaller failure strain at 300 K than 1.0 K for both the zigzag and armchair directions (see Fig. 7 (b)). Generally, ReaxFF gives a much better representation of the mechanical response of pristine black phosphorene over the SW potential.

## 3.4 Effect of Defects on the Mechanical Response of Black Phosphorene

Stress-strain curves of defected black phosphorene in the armchair and zigzag directions calculated with ReaxFF at 1 K are shown in Fig. 8 (a) and Fig. 8 (b), respectively. For armchair direction, black phosphorene with single vacancies shows a larger reduction in the failure strain than black phosphorene with double vacancies (keeping defect density the same), even though the double vacancy has a higher formation energy than the single vacancy. The reduction in the failure strain induced by Stone-Wales defect is in between that of single and double vacancy. The Young's modulus in the armchair direction is more or less unaffected by all three types of defects. For zigzag direction, all three types of defects reduce the failure strain by about 50%.

Only minor reduction in the Young's modulus in the zigzag direction is induced by all three types of defects. Thus, the mechanical response of black phosphorene in the zigzag direction is more sensitive to defects than that for the armchair direction.

To understand these phenomenon, the structural deformation and stress distribution of defected black phosphorene under tension ($\varepsilon$=0.13) in the armchair (Fig. 9 (a-c)) and zigzag (Fig. 9 (d-f)) directions were analyzed. For armchair direction (Fig. 9 (a-c)), stress at the single vacancy is more concentrated than that of double vacancy and Stone-Wales defect, due to the unsymmetrical defect geometry of single vacancy (double vacancy and Stone-Wales defect has central symmetry). Thus, the black phosphorene with single vacancies shows a larger reduction in the failure strain along the armchair direction than black phosphorene with double vacancies and Stone-Wales defects. Intriguingly, the structure of black phosphorene with single vacancy undergoes an unsymmetrical-to-asymmetrical transition induced by tension in the zigzag direction Fig. 9 (d). Consequently, three types of defects have similar influence on the mechanical response of black phosphorene under tension alone the zigzag direction.

Hao *et al.*[57] conducted first-principles study of the effect of single and double vacancies on the mechanical response of black phosphorene. The effect of single and double vacancies on the mechanical response of black phosphorene in both armchair and zigzag directions predicted by ReaxFF agrees fairly well with DFT results.[57] This clearly shows that ReaxFF for P/H systems provides a robust tool to study the effect of defects on the mechanical response of black phosphorene on a much larger space and time scale compared to DFT.

## 3.5 Thermal stability of phosphorene nanotubes

Similar to carbon nanotubes, the electrical and optical properties of the one-dimensional phosphorus nanotube (PNT) are chirality dependent and can be tuned by strain and size,[58–63] shedding light on its potential applications in transistors, strain sensors and photodetectors. Thus the accurate description of the properties of PNTs is important for ReaxFF. Two types of PNTs were designed by wrapping up a phosphorene sheet along the zigzag and armchair directions, i.e. (m, 0) zigzag PNTs and (0, n) armchair PNTs.[61,64] Fig. 10 shows that compared to SW potential,

ReaxFF provides a more accurate description of the cohesive energies change of the zigzag PNTs and armchair PNTs with respect to their sizes. SW potential underpredicts the cohesive energies of PNTs by an order of magnitude, indicating that SW potential could seriously underestimate the thermal stability of PNTs.[65] The phase diagrams for thermal stability of the zigzag PNTs and the armchair PNTs with varying temperatures and wrapping vectors of the nanotube are shown in Fig. 10 (c) and Fig. 10 (d), respectively. It is seen that SW potential strongly underpredicts the thermal stability of PNTs, compared to ReaxFF. Guan et al.[60] reported highly stable faceted PNTs can be constructed by laterally joining nanoribbons of different phosphorene phases. Intriguingly enough, ReaxFF for P/H is able to predict the phase transition of armchair and zigzag PNTs into faceted PNTs with higher thermal stability at elevated temperature, as shown in the inset figure. This discovery sheds light on the possible fabrication strategy of faceted PNTs. In short, ReaxFF is more reliable in describing the thermal stability of phosphorene nanotubes, compared to SW potential.

# 4. Concluding Remarks

We present a reactive force field (ReaxFF) for phosphorus and hydrogen, which gives an accurate description of the chemical and mechanical properties of pristine and defected black phosphorene. A 60° correction term is added which significantly improves the description of phosphorus clusters. ReaxFF for P/H is transferable to a wide range of P/H systems including bulk black phosphorus, blue phosphorene, phosphorus clusters and phosphorus hydride molecules. Emphasis has been placed on obtaining a good description of mechanical response of black phosphorene with different types of defects. Compared to SW potential, ReaxFF for P/H systems provides a notable improvement in describing the cohesive energy, mechanical response of pristine and defected black phosphorene and the thermal stability of phosphorene nanotubes. We observe a counterintuitive phenomenon that single vacancies weaken the black phosphorene more than relatively more unstable double vacancies. It is shown that the mechanical response of black phosphorene is more sensitive to defects in the zigzag direction than the armchair direction. Straightforward extensions to the heterogeneous systems, including oxides, nitrides, etc., enable the ReaxFF parameters for P/H systems to build a solid foundation for the simulation of a wide range of P-containing materials.

# Supporting Information

The ReaxFF file for P/H systems with or without 60 ° angle correction and the modified source file (*reaxc_valence_angles.cpp*) are available.

# Acknowledgments

X.C. acknowledges the support from the National Natural Science Foundation of China (11172231 and 11372241), ARPA-E (DE-AR0000396) and AFOSR (FA9550-12-1-0159). Many useful discussions with Mr. Mark Dittner are acknowledged.

# Figures

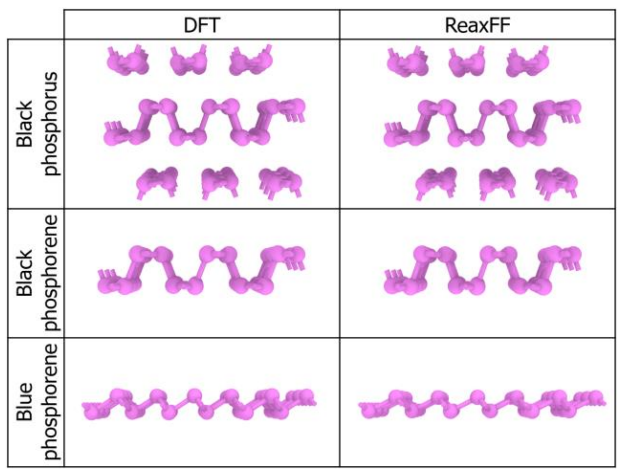

Fig. 1. Crystal structures of bulk black phosphorus, black phosphorene and blue phosphorene calculated by DFT and ReaxFF.

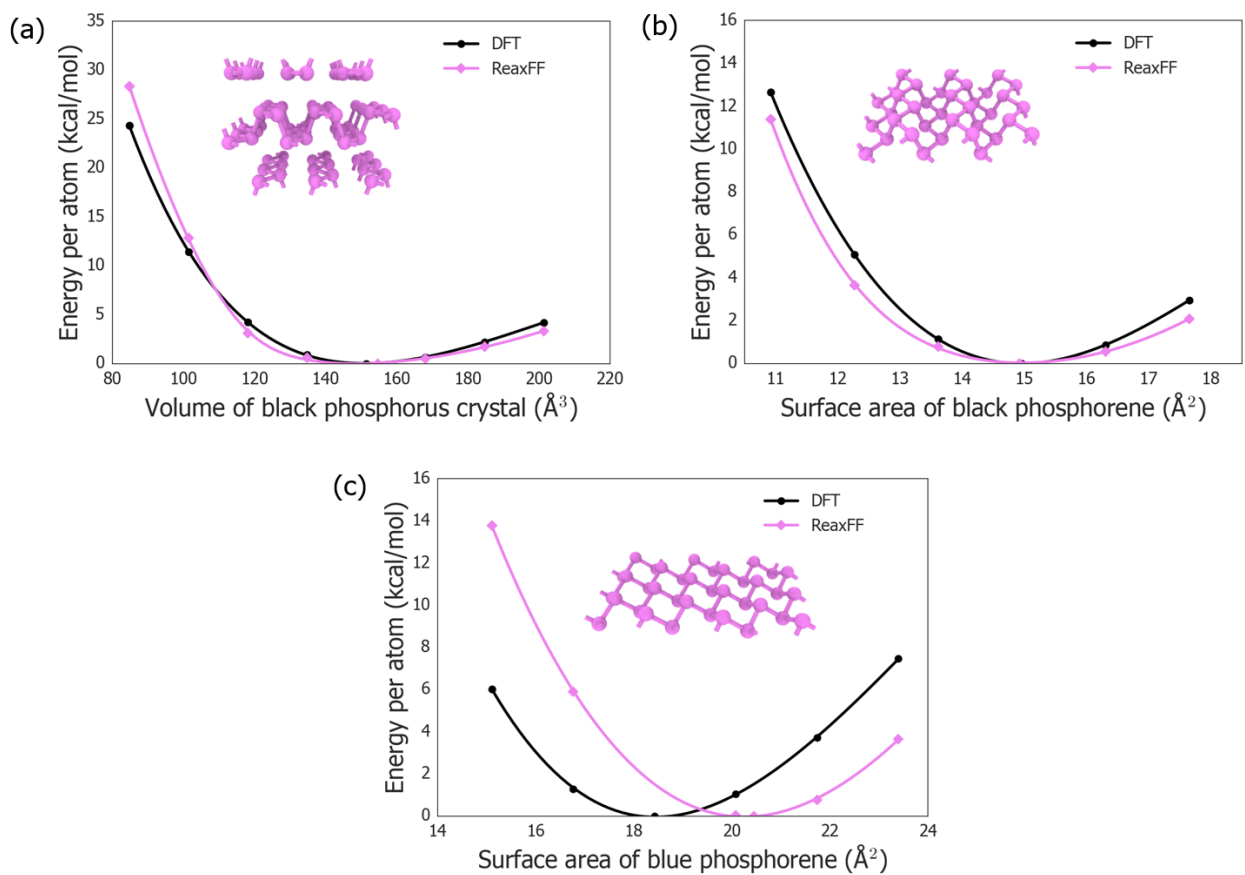

Fig. 2. Relative stabilities of (a) bulk black phosphorus for a broad range of unit cell volume, (b) black phosphorene for a broad range of in-plane unit cell area, (c) blue phosphorene for a broad range of in-plane unit cell area.

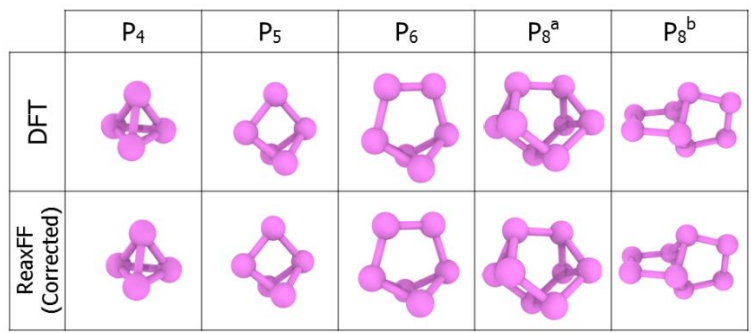

Fig. 3. Structures of phosphorus clusters from DFT and ReaxFF with the 60 ° correction.

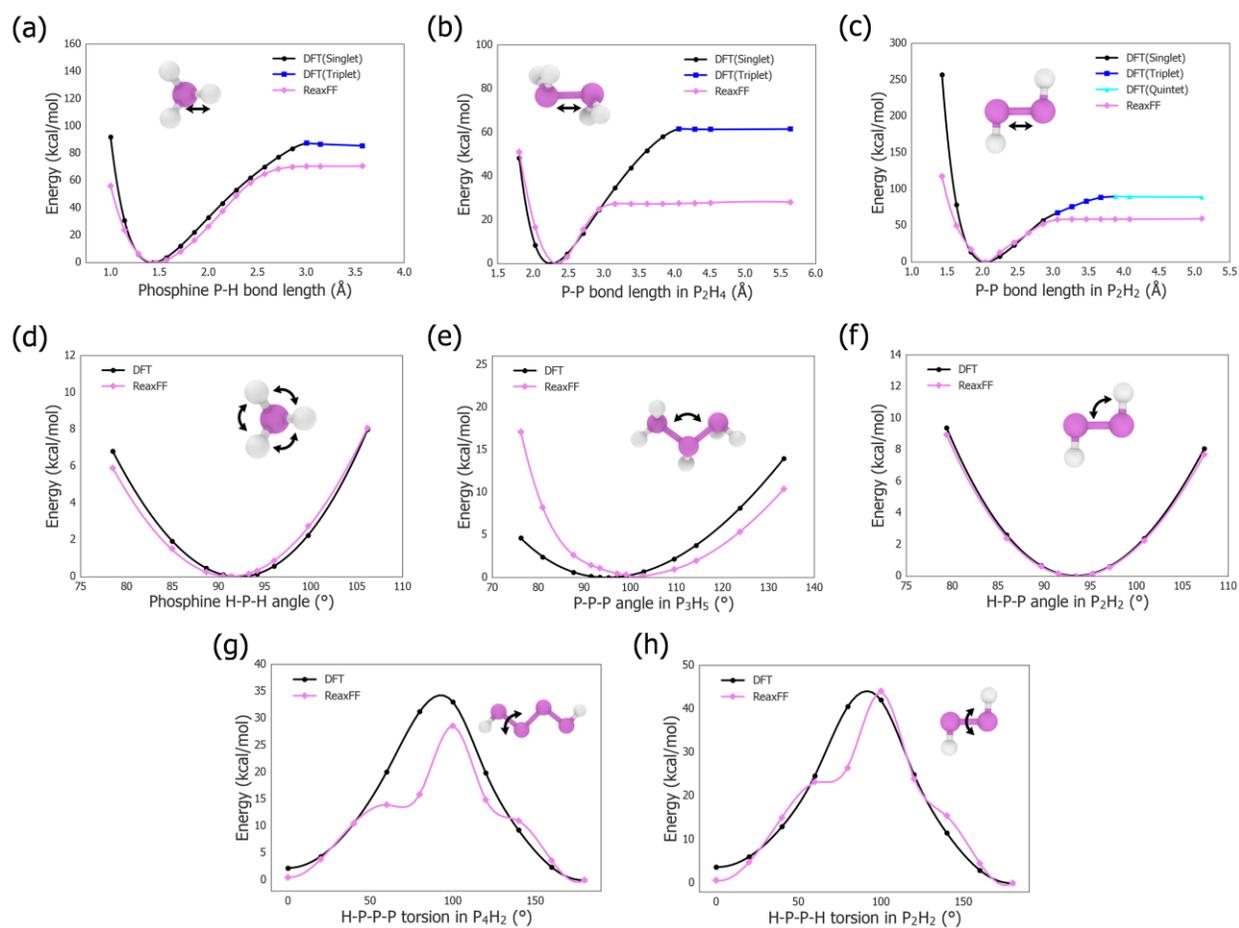

Fig. 4. DFT and ReaxFF potential energy curves for: (a) dissociation of a P-H bond in phosphine, (b) dissociation of a P-P bond in the $P_2H_4$ molecule, (c) dissociation of a P-P bond in the $P_2H_2$ molecule, (d) angle distortion of H-P-H in phosphine, (e) angle distortion of P-P-P in the $P_3H_5$ molecule, (f) angle distortion of H-P-P in the $P_2H_2$ molecule, (g) torsion distortion of H-P-P-H in the $P_2H_4$ molecule and of H-P-P-P in the $P_4H_2$ molecule.

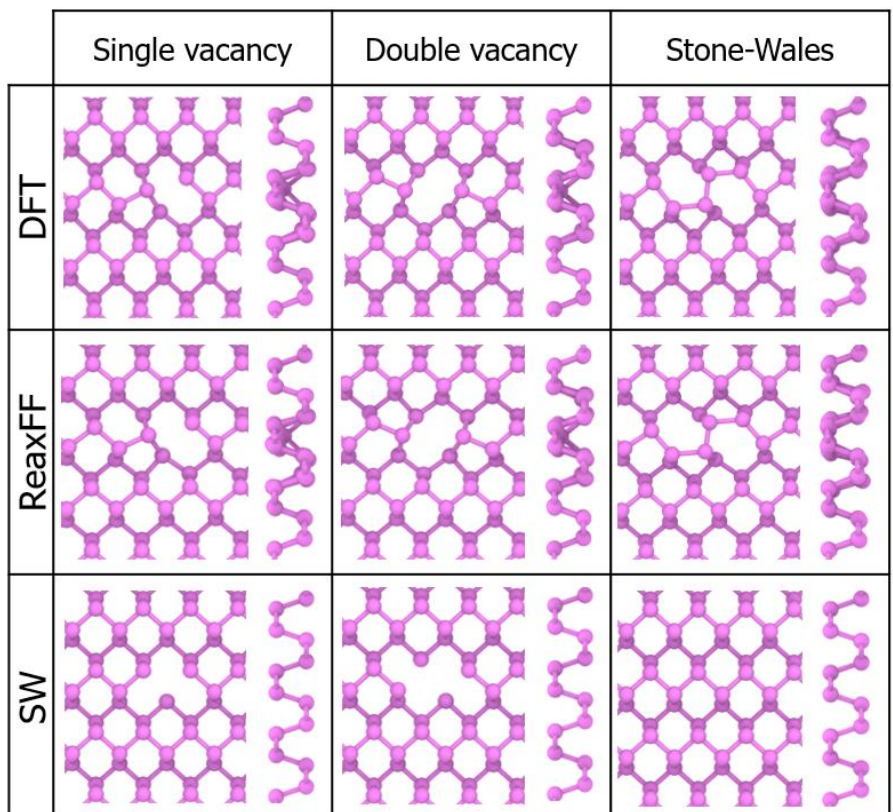

Fig. 5. Structures of defected black phosphorene calculated with DFT, ReaxFF and SW potential [1].

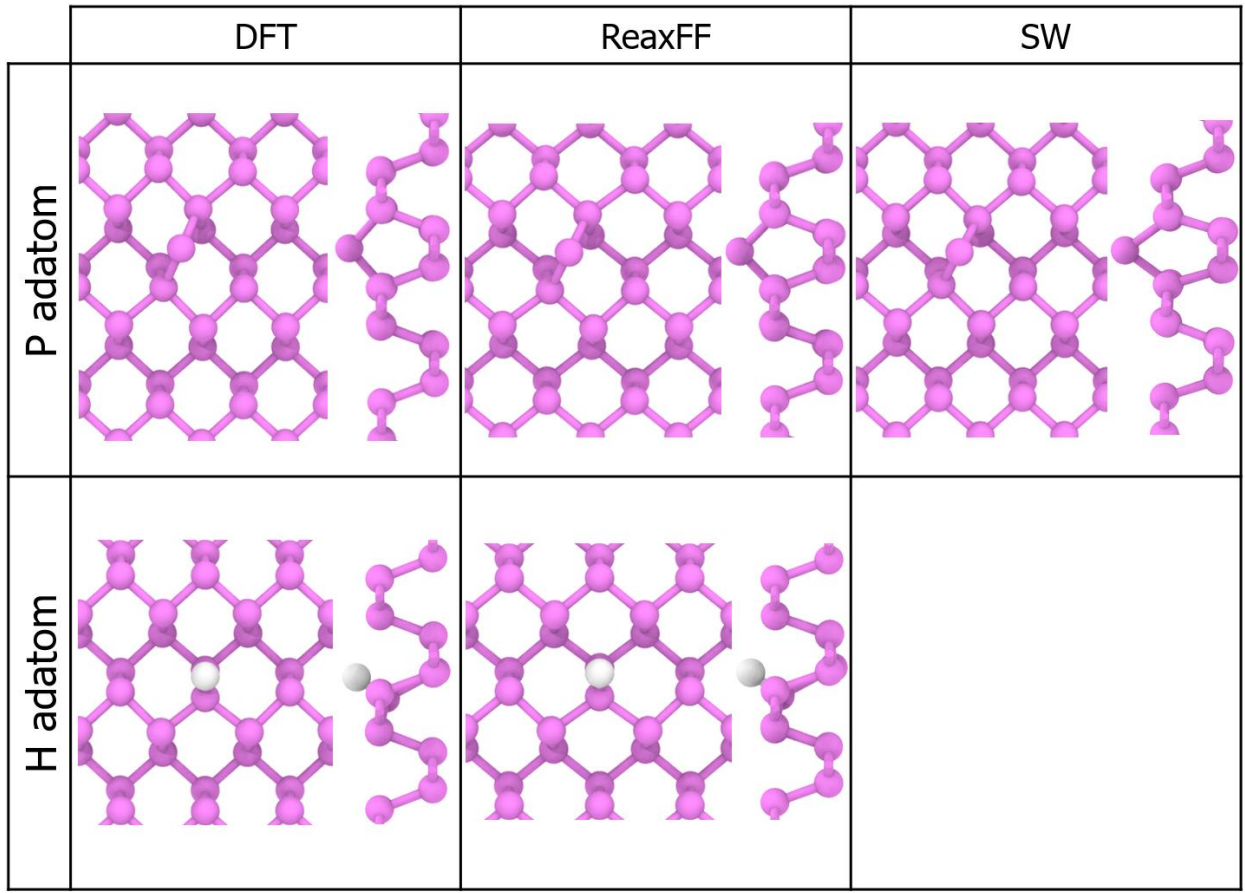

Fig. 6. Adsorption structures of P and H adatoms on black phosphorene calculated with DFT and ReaxFF compared to SW result (only for P adatom).

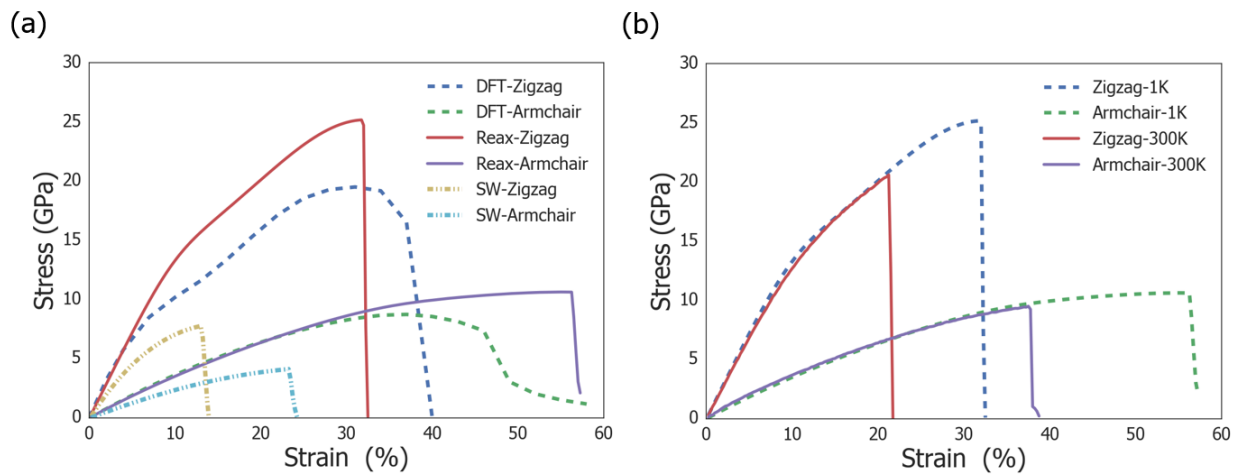

Fig. 7. (a) Stress-strain responses of black phosphorene along the armchair direction and zigzag direction calculated by ReaxFF and SW potential at 1 K compared to DFT results. (b) Stress-

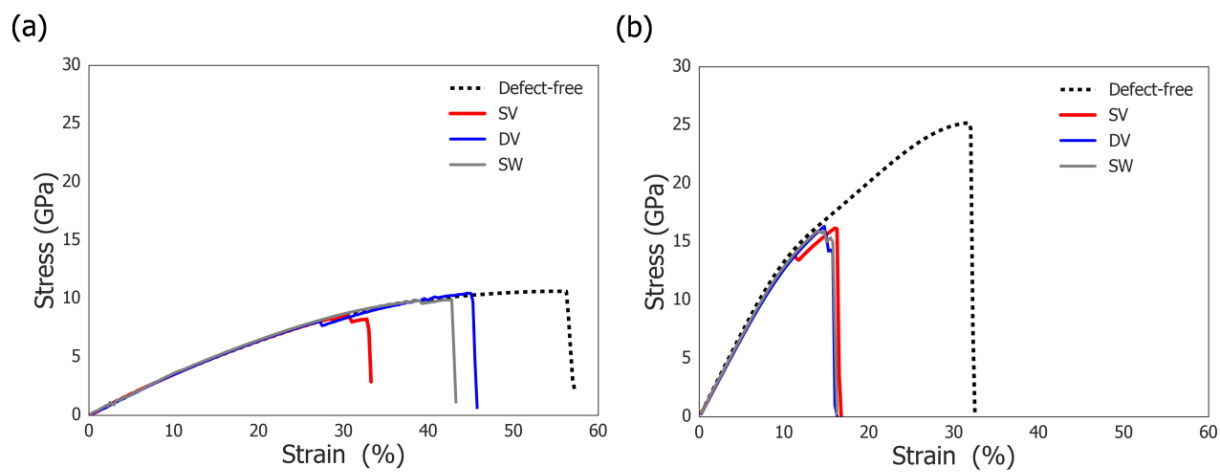

Fig. 8. Stress-strain responses of defected and defect-free black phosphorene along the armchair direction (a) and the zigzag direction (b) calculated by ReaxFF at 1K.

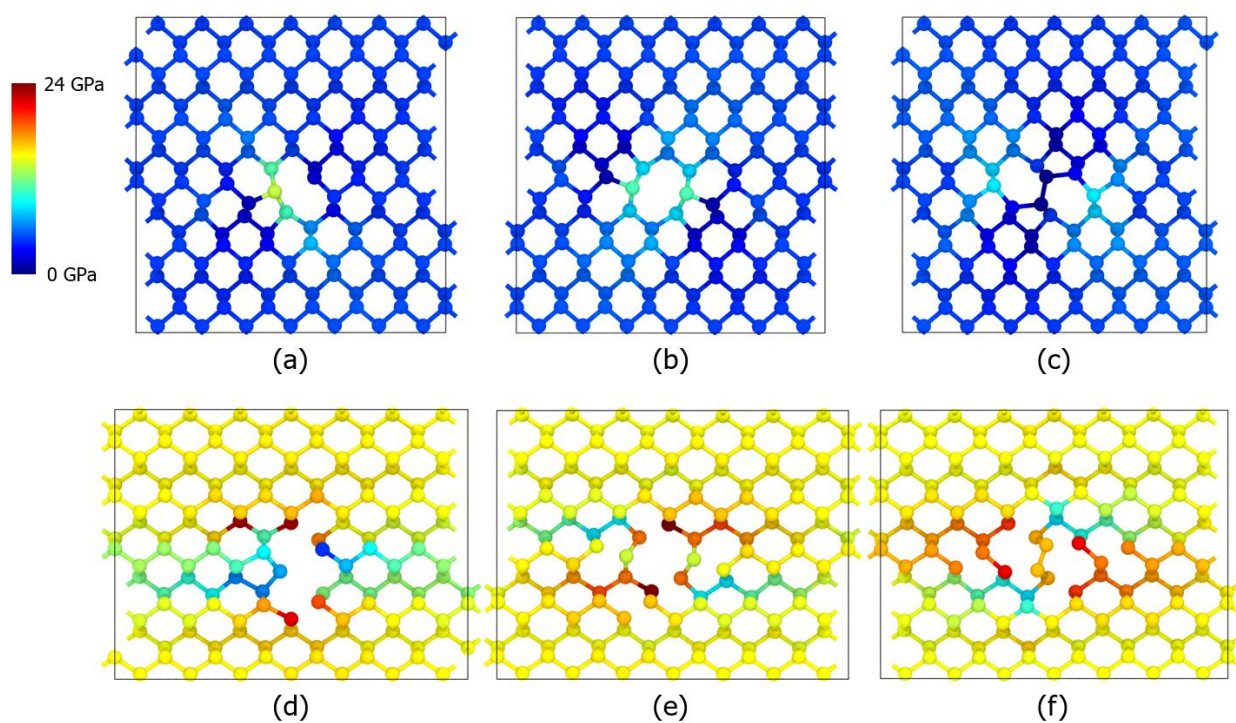

Fig. 9. Structure deformation and stress distribution of black phosphorene with single vacancy (a), double vacancy (b) and Stone-Wales defect (c) at $\varepsilon_{armchair} = 0.13$. Structure deformation and stress distribution of black phosphorene with single vacancy (d), double vacancy (e) and Stone-Wales defect (f) at $\varepsilon_{zigzag} = 0.13$. Colors show the stress distribution.

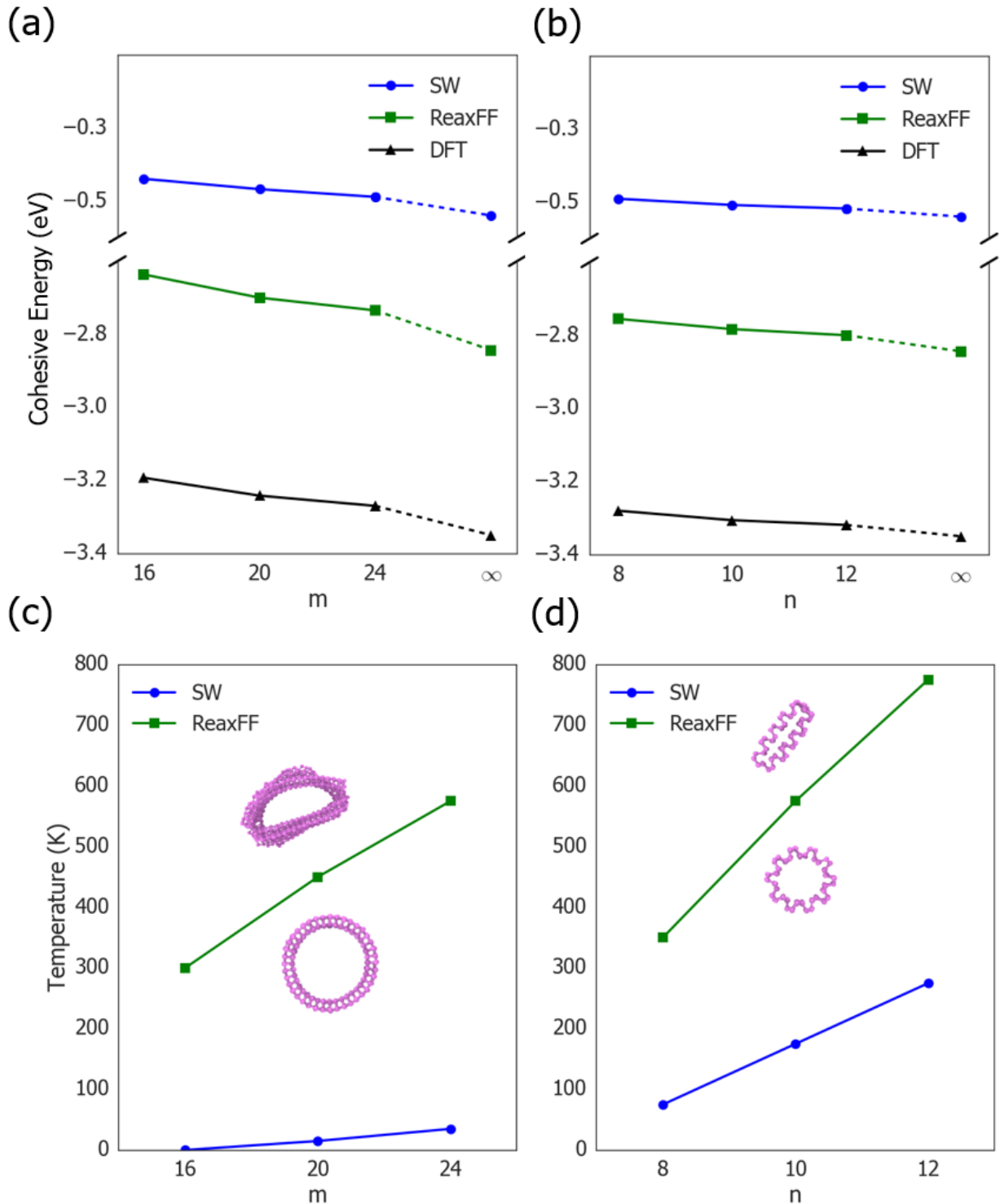

Fig. 10. Cohesive energies of the (m, 0) zigzag PNT (a) and (0, n) armchair PNT (b). The phase diagrams for thermal stability of the (m, 0) zigzag PNTs (c) and the (0, n) armchair PNTs (d) with varying temperatures and wrapping vectors of the nanotube. Stable and unstable PNT structures are shown.

# Tables

Table 1

Atom parameters for P and H

|   | Bond radii | | | $p_{ovun2}$ | Coulomb parameters | | | Bond order correction | | | Valence Angle | |
|---|---|---|---|---|---|---|---|---|---|---|---|---|
|   | $r_\sigma$ (Å) | $r_\pi$ (Å) | $r_{\pi\pi}$ (Å) |   | $\eta$ (eV) | $\chi$ (eV) | $\gamma$ (Å) | $p_{boc3}$ | $p_{boc4}$ | $p_{boc5}$ | $p_{val3}$ | $p_{val5}$ |
| P | 2.1199 | 1.9507 | 1.8354 | -2.0858 | 8.5658 | 6.3467 | 0.4060 | 15.5783 | 11.8556 | 2.8491 | 4.8954 | 1.6350 |
| H | 0.7853 |   |   | -15.7683 | 7.4366 | 5.3200 | 1.0206 | 3.3517 | 1.9771 | 0.7571 | 2.1488 | 2.8793 |

For H, parameters from Ref. [66] were used. Definitions of the individual ReaxFF parameters in this table and Tables 2–6 can be found in Refs. [26,27,31].

Table 2

vdW parameters and low-gradient vdW correction parameters for P and H

|   | van der Waals parameters | | | | lgvdW | | | | |
|---|---|---|---|---|---|---|---|---|---|
|   | $r_{vdW}$ (Å) | $\epsilon$ (kcal/mol) | $\alpha$ | $\gamma_{vdW}$ (Å) | $r_{core}$ (Å) | $\epsilon_{core}$ (kcal/mol) | $\alpha_{core}$ | $r_{lg}$ (Å) | $C_{lg}$ |
| P | 2.3355 | 0.0887 | 9.5120 | 7.6148 | 2.6552 | 0.0743 | 15.5028 | 2.1233 | 5066.5788 |
| H | 1.5904 | 0.0419 | 9.3557 | 5.0518 | 2.0000 | 0.0000 | 1.0000 | 1.9593 | 101.0453 |

For H, parameters from Ref. [66] were used.

Table 3

van der Waals and bond radius parameters for the P-H bond

|     | $r_\sigma$ (Å) | $r_{vdW}$ (Å) | $\epsilon$ (kcal/mol) | $\gamma_{vdW}$ (Å) |
|-----|------|------|------|------|
| P-H | 1.4319 | 1.5940 | 0.1064 | 10.3773 |

Table 4

Bond energy and bond-order parameters for the P-P, P-H and H-H bonds

| Bond | $D_e^\sigma$ (kcal/mol) | $D_e^\pi$ (kcal/mol) | $D_e^{\pi\pi}$ (kcal/mol) | $p_{be1}$ | $p_{be2}$ | $p_{ovun1}$ | $p_{bo1}$ | $p_{bo2}$ | $p_{bo3}$ | $p_{bo4}$ | $p_{bo5}$ | $p_{bo6}$ |
|------|------|------|------|------|------|------|------|------|------|------|------|------|
| P-P | 52.2711 | 23.4911 | 20.0346 | 0.4917 | 1.4218 | 0.7412 | -0.2457 | 7.5884 | -0.2226 | 13.6705 | -0.2395 | 17.8190 |
| P-H | 124.0512 | | | -0.3732 | 5.9712 | 0.5862 | -0.1003 | 5.6515 | | | | |
| H-H | 156.0973 | | | -0.1377 | 2.9907 | 0.8240 | -0.0593 | 4.8358 | | | | |

For the H-H bond, parameters from Ref. [66] were used.

Table 5

Valence angle parameters.

| Valence angle | $\Theta_{00}$ (degree) | $k_a$ (kcal/mol) | $k_b$ (1/rad)$^2$ | $p_{v1}$ | $p_{v2}$ |
|------|------|------|------|------|------|
| P-P-P | 81.1291 | 81.4496 | 0.5055 | 0.1993 | 1.0534 |
| H-P-P | 87.7897 | 48.0234 | 1.1576 | 2.4234 | 1.6028 |
| H-P-H | 91.5071 | 16.1001 | 2.6120 | 0.5531 | 1.0740 |

Table 6

| 60° angle correction | $\Theta_{60}$ (degree) | $p_{cor1}$ (kcal/mol) | $p_{cor2}$ (1/rad)$^2$ | $p_{cor3}$ |
|------|------|------|------|------|
| P-P-P | 60 | 16.6700 | 150.0000 | 1.0534 |

The parameters of 60° angle correction for P-P-P are designed to improve the description of phosphorus clusters with ReaxFF, explained in section 2.2.

Table 7

Torsion angle parameters

| General parameters | | Torsion angle | $V_1$ | $V_2$ | $V_3$ | $p_{tor1}$ |
|---|---|---|---|---|---|---|
| $p_{tor2}$ | 9.6260 | | | | | |
| $p_{tor3}$ | 9.7452 | H-P-P-P | -0.0137 | 46.5023 | 0.7269 | -3.2753 |
| $p_{tor4}$ | 4.1021 | H-P-P-H | -0.1595 | 49.6094 | 0.5875 | -2.0714 |

Table 8

DFT results and ReaxFF results (at 0 K) of bulk black phosphorus, black phosphorene and blue phosphorene compared to experimental obtained data.

| Structure | Lattice parameter | DFT (Å) | ReaxFF (Å) | Experiment (Å) |
|---|---|---|---|---|
| Bulk black phosphorus | $a$ | 3.30 | 3.46 | 3.31 [44] |
| | $b$ | 4.40 | 4.29 | 4.38 [44] |
| | $c$ | 10.43 | 10.43 | 10.48 [44] |
| Black phosphorene | $a$ | 3.28 | 3.46 | |
| | $b$ | 4.56 | 4.31 | |
| Blue phosphorene | $a$ | 3.26 | 3.43 | |
| | $b$ | 5.65 | 5.96 | |

Table 9

DFT results versus ReaxFF results of cohesive energies compared to SW results and experimental data

| Structure | Property | DFT | ReaxFF | SW | Experiment |
|---|---|---|---|---|---|
| Bulk black phosphorus | $E_{cohesive}$(bulk)/eV | -3.43 | -2.91 | | -3.26 [54] |

| | | | | |
|---|---|---|---|---|
| Black phosphorene | $E_{cohesive}$(black)/eV | -3.35 | -2.84 | -0.54 [1] |
| | $E_{cohesive}$(black) - $E_{cohesive}$(bulk) /(kcal/mol) | 1.94 | 1.58 | |
| Blue phosphorene | $E_{cohesive}$(blue) - $E_{cohesive}$(bulk) /(kcal/mol) | 3.00 | 2.15 | |

$E_{cohesive}$(bulk), $E_{cohesive}$(black) and $E_{cohesive}$(blue) are the cohesive energies of bulk black phosphorus, black phosphorene and blue phosphorene, respectively.

Table 10

Formation energy per atom of phosphorus clusters calculated by ReaxFF (with or without 60 ° correction) compared to DFT results.

| Cluster | Formation energy per atom (kcal/mol) | | |
|---|---|---|---|
| | DFT | ReaxFF | ReaxFF (60 ° correction) |
| $P_4$ | 7.6 | 56.9 | 7.6 |
| $P_5$ | 14.3 | 30.7 | 13.1 |
| $P_6$ | 11.3 | 25.1 | 9.7 |
| $P_8^a$ | 8.2 | 18.5 | 6.2 |
| $P_8^b$ | 12.7 | 11.1 | 11.8 |

Table 11

DFT results versus ReaxFF results of formation energies of SV, DV and SW defects in black phosphorene compared to SW results.

| Defect | Defect formation energy (eV) | | |
|---|---|---|---|
| | DFT | ReaxFF | SW [1] |
| SV | 1.66 | 1.80 | 0.54 |
| DV | 1.95 | 2.29 | 0.73 |
| SW | 1.42 | 1.94 | 0.00 |

Table 12

DFT results versus ReaxFF results of binding energies of phosphorus and hydrogen adatoms in black phosphorene compared to SW result (only for P).

| Atom | Adatom binding energy (eV) | | |
|---|---|---|---|
| | DFT | ReaxFF | SW [1] |
| P | -1.67 | -1.60 | -0.28 |
| H | -1.34 | -1.54 | |

Table 13

DFT results versus ReaxFF results of Young's modulus of black phosphorene in armchair and zigzag directions compared to SW results.

| | DFT | ReaxFF | SW [1] |
|---|---|---|---|
| $E_{arm}$ (GPa) | 37.8 | 38.4 | 33.5 |
| $E_{zig}$ (GPa) | 160.4 | 145.9 | 105.5 |
| $E_{zig} / E_{arm}$ | 4.24 | 3.81 | 3.15 |

## P-H ReaxFF parameters (without 60 degree correction):

```
Reactive MD-force field
 39       ! Number of general parameters
  50.0000 !Comment here
   9.5469 !Comment here
  26.5405 !Comment here
   1.7224 !Comment here
   6.8702 !Comment here
  60.4850 !Comment here
   1.0588 !Comment here
   4.6000 !Comment here
  12.1176 !Comment here
  13.3056 !Comment here
 -70.5044 !Comment here
   0.0000 !Comment here
  10.0000 !Comment here
   2.8793 !Comment here
  33.8667 !Comment here
   6.0891 !Comment here
   1.0563 !Comment here
   2.0384 !Comment here
   6.1431 !Comment here
   6.9290 !Comment here
   0.3842 !Comment here
   2.9294 !Comment here
  -2.4837 !Comment here
   9.6260 !Comment here
   9.7452 !Comment here
   4.1021 !Comment here
  -1.2327 !Comment here
   2.1645 !Comment here
   1.5591 !Comment here
   0.1000 !Comment here
   2.1365 !Comment here
   0.6991 !Comment here
  50.0000 !Comment here
   1.8512 !Comment here
   0.5000 !Comment here
   1.0000 !Comment here
   5.0000 !Comment here
   0.0000 !Comment here
   2.6962 !Comment here
  2    !Nr of atoms; cov.r; valency;a.m;Rvdw;Evdw;gammaEEM;cov.r2;
          alfa;gammavdW;valency;Eunder;Eover;chiEEM;etaEEM;n.u.
          cov r3;Elp;Heat inc.;n.u.;n.u.;n.u.;n.u.
          ov/un;val1;n.u.;val3,vval4
 P   2.1199  3.0000  30.9738  2.3355  0.0887  0.4060  1.9507  5.0000
     9.5120  7.6148  3.0000   0.0000 82.517   6.3467  8.5658  0.0000
     1.8354  0.0000 120.0000 11.8556 15.5783  2.8491  5066.5788  2.1233
    -2.0858  4.8954  1.0338   3.0000  1.6350  2.6552  0.0743 15.5028
  5066.5788  2.1233
 H   0.7853  1.0000  1.0080   1.5904  0.0419  1.0206 -0.1000  1.0000
     9.3557  5.0518  1.0000   0.0000 121.1250 5.3200  7.4366  1.0000
    -0.1000  0.0000 62.4879   1.9771  3.3517  0.7571  101.0453  1.9593
   -15.7683  2.1488  1.0338   1.0000  2.8793  2.0000  0.0000  1.0000
  101.0453  1.9593
   3     ! Nr of bonds; Edis1;LPpen;n.u.;pbe1;pbo5;13corr;pbo6
                       pbe2;pbo3;pbo4;n.u.;pbo1;pbo2;ovcorr
 1  1  52.2711 23.4911 20.0346  0.4917 -0.2395  1.0000 17.8190  0.7412
        1.4218 -0.2226 13.6705  1.0000 -0.2457  7.5884  1.0000  0.0000
 1  2 124.0512  0.0000  0.0000 -0.3732  0.0000  1.0000  6.0000  0.5862
        5.9712  1.0000  0.0000  1.0000 -0.1003  5.6515  0.0000  0.0000
 2  2 156.0973  0.0000  0.0000 -0.1377  0.0000  1.0000  6.0000  0.8240
        2.9907  1.0000  0.0000  1.0000 -0.0593  4.8358  0.0000  0.0000
   1    ! Nr of off-diagonal terms; Ediss;Ro;gamma;rsigma;rpi;rpi2
 1  2   0.1064  1.5940 10.3773  1.4319 -1.0000 -1.0000
   6    ! Nr of angles;at1;at2;at3;Thetao,o;ka;kb;pv1;pv2
 1  1  1  81.1291 81.4496  0.5055  0.0000  0.1993  0.000   1.0534
 2  2  2   0.0000  0.0000  5.8635  0.0000  0.0000  0.0000  1.0400
 2  1  1  87.7897 48.0234  1.1576  0.0000  2.4234  0.0000  1.6028
 2  1  2  91.5071 16.1001  2.6120  0.0000  0.5531  0.0000  1.0740
 1  2  1   7.0790  0.0000  0.4358  0.0000  0.0000  0.1050  2.1684
 2  2  1   0.0000  0.0000  6.0000  0.0000  0.0000  0.0000  1.0400
   2    ! Nr of torsions;at1;at2;at3;at4;;V1;V2;V3;V2(BO);vconj;n.u;n
 1  1  1  2 -0.0137 46.5023  0.7269 -3.2753  0.0000  0.0000  0.0000
 2  1  1  2 -0.1595 49.6094  0.5875 -2.0714  0.0000  0.0000  0.0000
   0    ! Nr of hydrogen bonds;at1;at2;at3;Rhb;Dehb;vhb1
```

## P-H ReaxFF parameters (with 60 degree correction):

```
Reactive MD-force field
 39       ! Number of general parameters
  50.0000 !Comment here
   9.5469 !Comment here
```

```
   26.5405 !Comment here
    1.7224 !Comment here
    6.8702 !Comment here
   60.4850 !Comment here
    1.0588 !Comment here
    4.6000 !Comment here
   12.1176 !Comment here
   13.3056 !Comment here
  -70.5044 !Comment here
    0.0000 !Comment here
   10.0000 !Comment here
    2.8793 !Comment here
   33.8667 !Comment here
    6.0891 !Comment here
    1.0563 !Comment here
    2.0384 !Comment here
    6.1431 !Comment here
    6.9290 !Comment here
    0.3842 !Comment here
    2.9294 !Comment here
   -2.4837 !Comment here
    9.6260 !Comment here
    9.7452 !Comment here
    4.1021 !Comment here
   -1.2327 !Comment here
    2.1645 !Comment here
    1.5591 !Comment here
    0.1000 !Comment here
    2.1365 !Comment here
    0.6991 !Comment here
   50.0000 !Comment here
    1.8512 !Comment here
    0.5000 !Comment here
    1.0000 !Comment here
    5.0000 !Comment here
    0.0000 !Comment here
    2.6962 !Comment here
  2    !Nr of atoms; cov.r; valency;a.m;Rvdw;Evdw;gammaEEM;cov.r2;
            alfa;gammavdW;valency;Eunder;Eover;chiEEM;etaEEM;n.u.
            cov r3;Elp;Heat inc.;n.u.;n.u.;n.u.;n.u.
            ov/un;val1;n.u.;val3,vval4
 P   2.1199   3.0000  30.9738   2.3355   0.0887   0.4060   1.9507   5.0000
     9.5120   7.6148   3.0000   0.0000  82.517   6.3467   8.5658   0.0000
     1.8354   0.0000 120.0000  11.8556  15.5783   2.8491           5066.5788   2.1233
    -2.0858   4.8954   1.0338   3.0000   1.6350   2.6552   0.0743  15.5028
 5066.5788   2.1233
 H   0.7853   1.0000   1.0080   1.5904   0.0419   1.0206  -0.1000   1.0000
     9.3557   5.0518   1.0000   0.0000 121.1250   5.3200   7.4366   1.0000
    -0.1000   0.0000  62.4879   1.9771   3.3517   0.7571 101.0453   1.9593
   -15.7683   2.1488   1.0338   1.0000   2.8793   2.0000   0.0000   1.0000
 101.0453   1.9593
  3       ! Nr of bonds; Edis1;LPpen;n.u.;pbe1;pbo5;13corr;pbo6
                        pbe2;pbo3;pbo4;n.u.;pbo1;pbo2;ovcorr
  1  1  52.2711  23.4911  20.0346   0.4917  -0.2395   1.0000  17.8190   0.7412
          1.4218  -0.2226  13.6705   1.0000  -0.2457   7.5884   1.0000   0.0000
  1  2 124.0512   0.0000   0.0000  -0.3732   0.0000   1.0000   6.0000   0.5862
          5.9712   1.0000   0.0000   1.0000  -0.1003   5.6515   0.0000   0.0000
  2  2 156.0973   0.0000   0.0000  -0.1377   0.0000   1.0000   6.0000   0.8240
          2.9907   1.0000   0.0000   1.0000  -0.0593   4.8358   0.0000   0.0000
  1       ! Nr of off-diagonal terms; Ediss;Ro;gamma;rsigma;rpi;rpi2
  1  2   0.1064   1.5940  10.3773   1.4319  -1.0000  -1.0000
  7       ! Nr of angles;at1;at2;at3;Thetao,o;ka;kb;pv1;pv2
  1  1  1  60.0000  16.6700 150.000   0.0000   0.1993   0.0000   1.0534
  1  1  1  81.1291  81.4496   0.5055   0.0000   0.1993   0.000    1.0534
  2  2  2   0.0000   0.0000   5.8635   0.0000   0.0000   0.0000   1.0400
  2  1  1  87.7897  48.0234   1.1576   0.0000   2.4234   0.0000   1.6028
  2  1  2  91.5071  16.1001   2.6120   0.0000   0.5531   0.0000   1.0740
  1  2  1   7.0790   0.0000   0.4358   0.0000   0.0000   0.1050   2.1684
  2  2  1   0.0000   0.0000   6.0000   0.0000   0.0000   0.0000   1.0400
  2       ! Nr of torsions;at1;at2;at3;at4;;V1;V2;V3;V2(BO);vconj;n.u;n
  1  1  1  2  -0.0137  46.5023   0.7269  -3.2753   0.0000   0.0000   0.0000
  2  1  1  2  -0.1595  49.6094   0.5875  -2.0714   0.0000   0.0000   0.0000
  0       ! Nr of hydrogen bonds;at1;at2;at3;Rhb;Dehb;vhb1
```